\documentclass[aps,prb,twocolumn,showpacs,superscriptaddress,reprint]{revtex4-1}
\usepackage{graphics}
\usepackage{xcolor}   
\usepackage{amsmath}
\usepackage{epsfig}
\usepackage{verbatim}

\usepackage[utf8]{inputenc}

\begin{document}
\title{Anharmonic quantum thermal transport across a van der Waals interface}
\author{Hangbo Zhou}
\affiliation{Institute of High Performance Computing, A*STAR, 138632, Singapore}
\author{Gang Zhang}
\email[]{zhangg@ihpc.a-star.edu.sg}
\affiliation{Institute of High Performance Computing, A*STAR, 138632, Singapore}
\author{Jian-Sheng Wang}
\affiliation{Department of Physics, National University of Singapore, 117551, Singapore}
\author{Yong-Wei Zhang}
\email[]{zhangyw@ihpc.a-star.edu.sg}
\affiliation{Institute of High Performance Computing, A*STAR, 138632, Singapore}
\date{\today}

\begin{abstract}
We investigate the anharmonic phonon scattering across a weakly interacting interface by developing a quantum mechanics-based theory. We find that the contribution from anharmonic three-phonon scatterings to interfacial thermal conductance can be cast into Landauer formula with transmission function being temperature-dependent. Surprisingly, in the weak coupling limit, the transmission due to anharmonic phonon scattering is unbounded with increasing temperature, which is physically impossible for two-phonon processes. We further reveal that the anharmonic contribution in a real heterogeneous interface (e.g., between graphene and monolayer molybdenum disulfide) can dominate over the harmonic process even at room temperature, highlighting the important role of anharmonicity in weakly interacting heterogeneous systems. 
\end{abstract}
\maketitle

\section{Introduction}
With the fast development in the discovery and synthesis of two-dimensional (2D) materials, van der Waals (vdW) heterostructures that are built by vertically stacking different 2D materials have drawn great interests in recent years \cite{Geim2013, Novoselov2016}. These vdW heterostructures not only serve as a new platform for exploring new physical phenomena, but also open up enormous possibilities for engineering applications, such as in photovoltaics \cite{Roy2013}, plasma devices \cite{Woessner2015} and field-effect tunnelling transistors \cite{Britnell947}. For example, a heterostructure formed by a monolayer graphene (Gr) and a monolayer molybdenum disulfide (MoS$_2$) has been experimentally constructed \cite{Azizi2015} and studied for applications in energy storage \cite{Chang2011, David2014, Wang2019}, water splitting \cite{Li2011} and photo-responsive devices \cite{Britnell2013, Roy2013}. Also with the rapid development of novel electronic devices based on 2D vdW heterostructures, their thermal management and heat dissipation are becoming increasingly challenging. In principle, the ultrahigh in-plane thermal conductivity of graphene could act as a heat dissipation media. However, since each membrane in such a weakly interacting heterostructure acts both as a bulk and as an interface, such construct can greatly limit the thermal conduction across the layers \cite{Liu2015, Ding2016}. In order to manage heat dissipation in such systems efficiently, an in-depth understanding and prediction of their interlayer thermal transport is necessary \cite{Tielrooij2018, Alborzi2020}. 

In many 2D materials, such as graphene and monolayer MoS$_2$, phonon is the dominating heat transport carrier. Due to the weak interactions and the mismatch of phonon dispersion between stacks in these vdW heterostructures, it is expected that inter-layer phonon-phonon interactions should be the limiting factor in the thermal management in such systems \cite{Gordiz2015, Le2017, Zhou2017}. Although the role of phonon anharmonicity in thermal transport has been studied extensively by using experiments and classical theories \cite{Tian2018, Zhou2010, Gordiz2015, Le2017, Zhou2017}, its quantum mechanical treatment is still a significant challenge. In dealing with bulk materials, many studies rely on the semi-classical Boltzmann transport equation by leveraging on approximated phonon relaxation times \cite{ShengBTE2014, Peng2016, Hu2020}. However, the breakdown of periodicity at an interface restricts the use of such treatment. In dealing with the bath-system-bath nano-junctions, the non-equilibrium Green's function method provides an alternative approach to treat phonon-phonon scattering through the calculation of high-order Feynman diagrams \cite{Mingo2006, Wang2007, Zhang2007, Xu2008, Dai2020}. However, the difficulties in calculating the high-order diagrams make such method in handling anharmonic scatterings a daunting task. To our knowledge, a regular quantum mechanical treatment on the cross-plane anharmonic phonon transport in vdW heterostructure is still lacking. As a result, many important questions that are associated with phonon anharmonicity in such weakly interacting systems are not answered. 
For example, though the well-known Meir-Wingreen formalism provides a general framework to calculate thermal conductance due to anharmonic phonons, the temperature dependence of thermal conductance is not explicit, due to the unsolvable nonlinear self-energy. However, with the condition of the weak coupling limit, it becomes possible to look into its temperature dependence. 
As another example, practically how important the interlayer anharmonic phonon scattering is compared with its harmonic counterpart at room temperature is not clear. Clearly, answers to these questions are not only of great scientific interest in term of revealing new thermal physics in weakly interacting heterostructures, but also of important technological impact to the thermal management and heat dissipation in such heterostructures.  

To examine the importance of phonon anharmonicity and understand the thermal inter-layer phonon scattering mechanism across vdW heterostructure, here, we focus on the two-phonon (harmonic) and three-phonon (anharmonic) scattering processes. To handle the anharmonic phonon transport across a weakly interacting interface, we develop a quantum transport theory, which places both harmonic scattering and anharmonic scattering under the same theoretical framework.  By using this theory, we investigate the phonon transmission across a weakly interacting interface, aiming to answer the above-mentioned questions.   

\section{Theoretical framework}
\subsection{Phenomenological description}

\begin{figure}
    \centering
    \includegraphics[width=\linewidth]{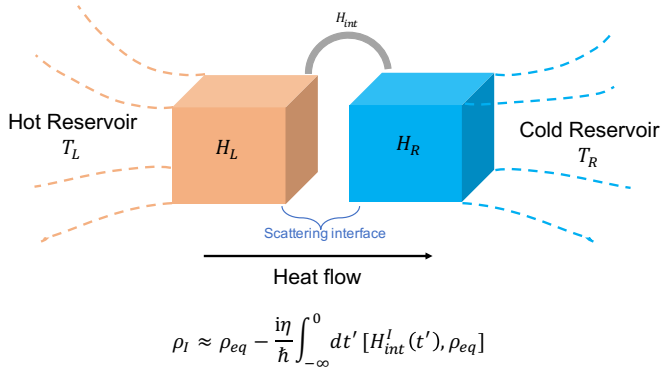}
    \caption{Schematic for the physical model. The interface model comprises a left lead and a right lead. The leads are weakly interacting with each other. Each lead is coupled to a respective thermal reservoir. The reservoirs are assumed to be in equilibrium, and they are assigned with temperature $T_L$ and $T_R$ for the left and right reservoir respectively.}
    \label{fig:Model}
\end{figure}

Our model comprises a hot reservoir ($H_H$), a cold reservoir ($H_C$), a left lead ($H_L$), a right lead ($H_R$), an interface between the hot reservoir and left lead ($H_{HL}$),  an interface between the right lead and cold reservoir ($H_{RC}$) and an interface between the left and right lead ($H_{int}$), as shown in Fig.~\ref{fig:Model}. So the entire system will evolve under the total Hamiltonian,
\begin{equation}
    H_{tot}=H_H+H_{HL}+H_L+\eta H_{int}+H_R+H_{RC}+H_C
\end{equation}
Here, we assume that the interaction between the left and right leads is weak in comparison with all other interactions. We use a small quantity $\eta$ to characterize that weak coupling.
The leads are modelled by a collection of phonons so they are governed by the left-lead Hamiltonian ($H_L=\sum_q[\frac{(\tilde{p}^L_q)^2}{2m}+\frac{1}{2}\omega_q^2(\tilde{x}_q^L)^2]$) and right-lead Hamiltonian  ($H_R=\sum_q[\frac{(\tilde{p}^R_q)^2}{2m}+\frac{1}{2}\omega_q^2(\tilde{x}_q^R)^2]$). The Hamiltonians of reservoirs are often complicated and in many cases they are unknown. The effect of the reservoirs is to continuously thermalize the leads by injecting or draining phonons, while the lead-lead coupling is to drive each of the lead out-of-equilibrium. Since we focus on the lead-lead interfacial resistance, we assume that the thermal resistance due to reservoir-lead coupling is negligible compared to the lead-lead interfacial resistance. 
So the bottleneck of a travelling phonon is at the lead-lead interface. Thus in this section we phenomenologically analyze possible routines of phonons travelling across the lead-lead interface through two-phonon and three-phonon processes. Detailed theoretical derivations will be presented in the next section.

Suppose the interfacial coupling is governed by an atomic potential function $V(\vec{r}^L,\vec{r}^R)$, then the Hamiltonian of atomistic coupling across the interface up to the third-order can be written as
\begin{eqnarray}
    \label{eq:ifc}
    \eta H_{int}&=&\sum_{i\in L,j\in R}\frac{1}{2}K_{ij}x^L_ix^R_j+\sum_{i\in L,jk\in R}\frac{1}{6}V_{i,jk}x^L_ix^R_jx^R_k\nonumber\\
    &+&\sum_{ij\in L,k\in R}\frac{1}{6}V_{ij,k}x^L_ix^L_jx^R_k,
\end{eqnarray}
where $x^L_i=dr_i^L$ is the displacement of atomic degree of freedom $i$ from its equilibrium position at part $L$. A similar notation also applies to $R$. $K_{ij}=\frac{\partial V(\vec{r}^L,\vec{r}^R)}{\partial r^L_i\partial r^R_j}$ are the second-order interatomic force constants, which govern the two-phonon processes. $V_{i,jk}=\frac{\partial V(\vec{r}^L,\vec{r}^R)}{\partial r^L_i\partial r^R_j\partial r^R_k}$ and $V_{ij,k}=\frac{\partial V(\vec{r}^L,\vec{r}^R)}{\partial r^L_i\partial r^L_j\partial r^R_k}$ are the third-order interatomic force constants, which govern the three-phonon processes. For the sub-indices of $V$, we use the notation that the indices before the comma denote the degrees of freedom of the left bath and indices after the comma denote that of the right bath. Here the interatomic force constants $K$ and $V$ are small and inherited from $\eta$. The displacement, for example $x^L_i$, can be expanded with respect to the displacement of the normal modes of phonons with wave vector $q$ in $L$ as $x^L_i=\sum_q c_i^q \tilde{x}^L_q$. 

The displacement operator $\tilde{x}_q=\sqrt{\frac{\hbar}{2m\omega_q}}(a_q^\dagger+a_q)$ is quantized with $a^\dagger_q$ and $a_q$  being creation and annihilation operations of phonon with wave vector $q$.
It is now obvious that the second-order term in Eq.(\ref{eq:ifc}) governs the process that a phonon is annihilated at part $L$ and meanwhile a phonon is created at part $R$, or vice versa. The two phonons should have the same frequency due to energy conservation requirement but not necessarily the same wave vector. It is noted that a two-phonon process implies elastic scattering. However, the scattering processes due to the third-order terms become much more complex. In this case, the frequency of annihilated phonon(s) differs from the created phonon(s) because the total number of phonons is not conserved. The energy conservation only guarantees that the total summation of frequencies of annihilated phonons equals the summation of frequencies of created phonons. 

\begin{figure}
    \centering
    \includegraphics[width=0.8\linewidth]{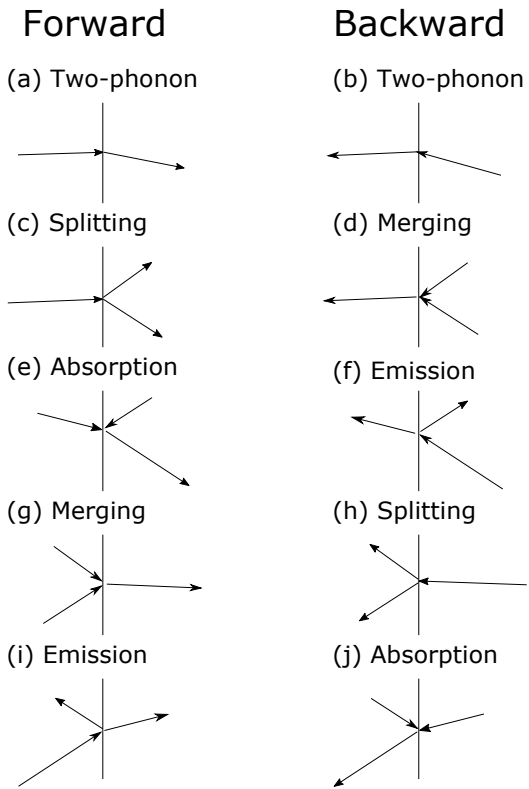}
    \caption{Two-phonon scattering processes (a and b) and three-phonon scattering processes (c-j).}
    \label{fig:scattering}
\end{figure}

The detailed scattering processes due to two-phonon interactions and three-phonon interactions are illustrated in Fig.~\ref{fig:scattering}. These scattering processes are determined by the phonon density of states of both sides, the interface coupling details and the temperature of each side. We first discuss the effect of temperature. Temperature will determine the phonon occupation number of a particular phonon state. Therefore, given two or three phonon states at two sides of the interface, the phonons can undergo both forward scattering (such (a), (c), (e), (g) and (i)) and backward scattering (such as (b), (d), (f), (h), and (j)). If the temperature of the left region is greater than the temperature of the right region ($T_L>T_R$), the forward processes will dominate over the backward processes, resulting in a net heat flow from L to R. In thermal equilibrium ($T_L=T_R$), the forward and corresponding backward processes (such as (c) and (d)) will have the same occurrence probability. Therefore, the overall net heat flow is zero.

Besides temperature, the occurrence probability of each scattering process also depends on the material at each side (which decides the spectrum of phonon frequencies at each side) and the details of interface coupling (which decides the number of phonons participating in the scatterings at each side). If both sides have the same material and the coupling is also symmetric, then the forward scattering process will also be symmetric with the backward scattering process. For example, if the forward splitting process (a) is symmetric with the backward splitting process (f), or the forward absorption process (e) is also symmetric with the backward absorption process (j), the thermal conductance of forward and backward transports will be the same and hence, the thermal rectification effect should disappear.

\subsection{Theoretical analysis}

The above phenomenological analysis provides a basic understanding for the phonon scattering processes at the interface. However, many important insights can only be revealed by using formal anharmonic quantum transport theory. For example, the underlying reason why two-phonon scattering process cannot generate thermal rectification effect even at a heterogeneous interface only becomes obvious after the Landauer transport theory was established for elastic phonon scatterings. More specifically, the Landauer theory can be cast as:
\begin{equation}
    I_h=\int_0^\infty \frac{d\omega}{2\pi}\hbar\omega \bar{T}(\omega)[n_L(\omega)-n_R(\omega)],
\end{equation}
where $I_h$ is the thermal current flowing out of the lead region. Here $\bar{T}(\omega)$ is the temperature-independent transmission function of phonons at frequency $\omega$. (We used the notation $\bar{T}$ to distinguish the transmission from temperature $T$.). $n_L$($n_R$) is the Bose-Einstein distribution of phonons in heat bath $L(R)$. Landauer formula also reveals the temperature-dependence of thermal conductance and the presence of a saturated thermal conductance at a high temperature.

Compared to two-phonon scattering processes, much less is known about three-phonon scattering processes. Although thermal rectification effect due to three-phonon scatterings (the difference in thermal conductance upon temperature reversal) has been reported, the origin of the rectification effect and the necessary condition that such rectification exists are still not fully understood. Intuitively, due to the requirement of a larger number of phonons involved, the thermal conductance due to three-phonon scatterings should be minor at low-temperature, but become increasingly important with the increase of temperature. But for three-phonon processes, whether there is a saturation in the thermal conductance at a high temperature, which is the case for two-phonon processes, is not clear.  

In this work, we explore the thermal current in the weak coupling limit. We separate the total system into two subsystems, that is, the two reservoirs and the rest consisting of the two leads and their interfaces. We denote the total density matrix as $\varrho$ and the density matrix of the leads as [$\rho=Tr_{H,C}(\varrho)$]. We can then prove that the evolution of the density matrix of the leads is governed by 
\begin{equation}
\frac{d\rho(t)}{dt}=-\frac{i}{\hbar} [H_L+H_R+\eta H_{int},\rho(t)]+ \mathcal{D}(\rho),
\label{eq:QME}
\end{equation}
where the first term is the self-evolution of the leads, and the second term describes the dissipation and thermalization effects from the two reservoirs.  We note that, due to the weakness of the interfacial coupling, the time scale for a phonon to cross the lead-reservoir interface is much shorter than the time scale for a phonon to cross the lead-lead interface.

We adopted the adiabatic switch-on techniques such that the left and right regions are decoupled at $t=-\infty$ and they are at their own thermal equilibrium with its reservoir and possesses a density matrix of $\rho_{eq}=\rho_L\otimes\rho_R\propto e^{-\beta_L H_L}\otimes e^{-\beta_R H_R}$, with $\beta_\alpha=1/(k_B T_\alpha), \alpha=L,R$ being inverse temperatures.  Subsequently, their interface coupling is adiabatically turned on. The leads are  perturbed and driven out of equilibrium by the temperature difference between the two leads. However, due to the fast thermalization of the lead by the reservoir, such weak interfacial coupling between the two leads only causes a small perturbation and the lead remains in a near equilibrium state with its own reservoir. With this assumption, Eq.(\ref{eq:QME}) can be approximated by a linear response of the lead-lead interface coupling $H_{int}$ to the equilibrium state:
\begin{equation}
\rho(t)=\rho_{eq}-\frac{i\eta}{\hbar} \int_{-\infty}^0 [H_L+H_R+\eta H_{int},\rho_{eq}(t)]  ,
\end{equation}
By transforming into interaction picture, this equation becomes
\begin{equation}
\label{eq:rho}
    \rho_I \approx \rho_{eq}-\frac{i\eta}{\hbar}\int_{-\infty}^0 dt'[H_{int}^I(t'),\rho_{eq}],
\end{equation}
where $\rho_I(t)=e^{iH_0 t/\hbar}\rho(t) e^{-iH_0/\hbar}$ and $H^I_{int}(t)=e^{iH_0 t/\hbar}H_{int}e^{-iH_0/\hbar}$ with $H_0=H_L+H_R$. 
The first term is the equilibrium state density matrix and the second term represents the non-equilibrium states caused by the lead-lead interface coupling.

Equation (\ref{eq:rho}) provides a method to evaluate non-equilibrium quantum states $\rho_I$. In order to obtain the thermal current, we need a thermal current operator, which can be defined as  $I_L=-d(H_H+H_{HL}+H_L)/dt$. We assume that the left reservoir and the reservoir-lead coupling is not interacting with the lead-lead coupling: $[H_H, H_{int}]=0$ and $[H_{HL}, H_{int}]=0$. If this is not the case, we may re-partition the reservoir and leads such that all the degrees of freedom involving in the interface coupling should be comprised within the leads. With this assumption the current observable can be simplified to
\begin{equation}
    I_L=-d(H_H+H_{HL}+H_L)/dt=-\frac{i\eta}{\hbar}[H_{int},H_L]
\end{equation}

By plugging the above expansion into the current expression $I_h=\mbox{Tr}(\rho_I I_L)$, at $t=0$, we find that the thermal current can be expanded according to the coupling strength $\eta$ as
\begin{equation}
    I_{h}=a_1\eta+a_2\eta^2+a_3\eta^3+\cdots,
\end{equation}
where the first-order term vanishes ($a_1=-\frac{i}{\hbar}\mbox{Tr}(\rho_{eq}[H_{int},H_L])=0$) due to the fact $[\rho_{eq}, H_L]=0$. This is consistent with the literature\cite{Thingna2012, Thingna2014, Zhou2020} that the lowest order current arises from the second-order of the coupling strength $\eta$. 
As a result, the lowest order term comes from $a_2\eta^2$, which is
\begin{equation}
    a_2\eta^2=-\frac{\eta^2}{\hbar^2}\int_{-\infty}^0dt'\mbox{Tr}\Big([H_{int}^I(t'),\rho_{eq}][H_{int},H_L]\Big)
\end{equation}

Due to the fact that $H_{int}$ contains second- and third-order interactions, we find that $a_2\eta^2$ comprises contributions from two-phonon process and three-phonon process $a_2\eta^2=I_{2p}+I_{3p}$. By plugging in the explicit form of $H_{int}$, as shown in Eq.(\ref{eq:ifc}), a straightforward derivation will reveal that the thermal current due to two-phonon process is given by,\cite{Zhou2020} 
\begin{equation}
    I_{2p}=-\frac{1}{4\hbar}\sum_{ij,kl}K_{i,j}K_{k,l}\int_{-\infty}^\infty\Psi_{ik}(t)\Phi_{jl}(t)dt,
\end{equation}
where $\Psi_{ij}(t)=\frac{d\Phi_{ij}(t)}{dt}$, $\Phi_{ij}(t)=\left<x_i(t)x_j\right>$ are the two-point position correlation functions within each bath. Here $\langle\cdots\rangle$ stands for the evaluation of expectation under equilibrium density matrix $\mbox{Tr}(\rho_{eq}\cdots)$. The above term comes from the first term in Eq.~(\ref{eq:ifc}), which governs two-phonon forward and backward scattering processes (Fig.\ref{fig:scattering}(a) and \ref{fig:scattering}(b)). On the other hand, the contribution from the second term of Eq.~(\ref{eq:ifc}) is
\begin{equation}
\label{eq:3pha}
    I_{3p}^a=-\frac{i}{36\hbar}\sum_{ijk,lmn}V_{i,jk}V_{l,mn}\int_{-\infty}^\infty\Psi_{il}(t)\Phi_{jkmn}(t)dt,
\end{equation}
where $\Phi_{jkmn}(t)=\left<x_j(t)x_k(t)x_mx_n\right>$ is the four-point correlation function of side $R$.
 We can see that this term deals with one phonon at the left side and two phonons at the right side so that it describes the scattering processes of forward splitting (Fig.\ref{fig:scattering}(c)), backward merging (Fig.\ref{fig:scattering}(d)), forward absorption (Fig.\ref{fig:scattering}(e)), and backward emission (Fig.\ref{fig:scattering}(f)). Finally, the contribution from the last term of Eq.~(\ref{eq:ifc}) is
\begin{equation}
\label{eq:3phb}
    I_{3p}^b=-\frac{i}{36\hbar}\sum_{ijk,lmn}V_{ij,k}V_{lm,n}\int_{-\infty}^\infty\Psi_{ijlm}(t)\Phi_{kn}(t)dt,
\end{equation}
where $\Psi_{ijlm}(t)=\frac{d\Phi_{ijlm}(t)}{dt}$. Similarly, it deals with two phonons at the left part and one phonon at the right part so that it describes the scattering processes of forward merging (Fig.\ref{fig:scattering}(g)), backward splitting (Fig.\ref{fig:scattering}(h)), forward emission (Fig.\ref{fig:scattering}(i)), and backward absorption (Fig.\ref{fig:scattering}(j)). 

With the help of Eq.(\ref{eq:rho}), the non-equilibrium density matrix can be evaluated from the equilibrium product state. As a result, the correlation functions are evaluated under their own respective equilibrium states. Therefore Wick's theorem can be applied and these four-point correlation functions can be decoupled into the summations of two-point correlations. The results are $\Phi_{ijkl}(t)=\Phi_{ik}(t)\Phi_{jl}(t)+\Phi_{il}(t)\Phi_{jk}(t)$ and $ \Psi_{ijkl}(t)=\Psi_{ik}(t)\Phi_{jl}(t)+\Psi_{jl}(t)\Phi_{ik}(t)+\Psi_{il}(t)\Phi_{jk}(t)+\Psi_{jk}(t)\Phi_{il}(t)$. We omitted the term $\Phi_{ij}(0)\Phi_{kl}(0)$ since its contribution to thermal conductance is zero. The two-point correlation functions can be written in terms of the spectral densities of the left part $\Gamma_L$ and right part $\Gamma_R$ as $\Phi_{ij}(t)=\int_{-\infty}^\infty\frac{d\omega}{\pi}\Gamma_{ij}(\omega)n(\omega)e^{i\omega t}$ and $\Psi_{ij}(t)=i\int_{-\infty}^\infty\frac{d\omega}{\pi}\Gamma_{ij}(\omega)\omega n(\omega)e^{i\omega t}$. The spectra density is defined as
\begin{equation}
    \Gamma_{ij}(\omega)=\pi\sum_q\frac{\hbar c_i^q c_j^q}{2m\omega_q}\delta(\omega-\omega_q),
\end{equation}
where $\omega_q$ is the phonon dispersion relation. 
In terms of Green's function functions, this spectral density is actually $\Gamma_{ij}=\frac{i\hbar}{2}[g_{ij}^r(\omega)-g_{ij}^a(\omega)]$, where $g^r(\omega)$ and $g^a(\omega)$ are retarded and advanced surface Green's functions. 
Therefore, in practical calculation, we can first evaluate the surface Green's function by using a numerical scheme \cite{Wang2008} and then obtain the spectra density from the surface Green's functions.

Rearranging the above formulas, one can cast the total thermal current into Landauer formula. However, the marked difference is that the two-phonon processes give a temperature-independent transmission function, while the three-phonon processes give a temperature-dependent transmission function. 
Explicitly, we show that the lowest order current is
\begin{equation}
    I_h=\int_0^\infty \frac{d\omega}{2\pi}\hbar\omega [\bar{T}_{2ph}(\omega)+\bar{T}_{3ph}(\omega,T_L,T_R)][n_L(\omega)-n_R(\omega)].
\end{equation}
The two-phonon transmission function is $\bar{T}_{2p}(\omega)=K_{i,j}K_{k,l}\Gamma_{ik}(\omega)\Gamma_{jl}(\omega)$, while the three-phonon transmission function consists of two terms arising from Eq.~(\ref{eq:3pha}) and Eq.~(\ref{eq:3phb}), that is, $\bar{T}_{3p}(\omega,T_L,T_R)=\bar{T}_a(\omega,T_L,T_R)+\bar{T}_b(\omega,T_L,T_R)$, and 
\begin{equation}
    \bar{T}_\alpha(\omega,T_L,T_R)=-\frac{8}{\hbar^2}\int_{-\infty}^{\infty}\frac{d\omega'}{2\pi}S_\alpha(\omega,\omega')f_\alpha(T_L,T_R,\omega,\omega')
\end{equation}
where $\alpha(=\!\!\!\!a,b)$ is the term index, $f$ contains phonon distribution function and $S$ is the scattering matrix. Explicitly, they are
$S_a(\omega,\omega')=\frac{1}{36}\sum\limits_{il,jkmn} \allowbreak V_{i,jk}V_{l,mn}[\Gamma_{il}(\omega)\Gamma_{jm}(\omega'')\Gamma_{kn}(\omega')+\Gamma_{il}(\omega)\Gamma_{jn}(\omega'')\Gamma_{km}(\omega')]$ 
and
$S_b(\omega,\omega')=\frac{1}{36}\sum\limits_{ij,lmkn} \allowbreak V_{ij,k}\allowbreak V_{lm,n}\allowbreak [\Gamma_{il}(\omega')\Gamma_{jm}(\omega'')\Gamma_{kn}(\omega)+\Gamma_{im}(\omega')\Gamma_{jl}(\omega'')\Gamma_{kn}(\omega)]$, where $\omega''=-(\omega+\omega')$. The phonon distribution function turns out to be
\begin{equation}
\label{eq:Tdep1}
    f_a(\omega,\omega')=\frac{e^{\beta_R\hbar\omega'}(e^{\beta_R\hbar\omega}-1)}{(e^{\beta_R\hbar(\omega+\omega')}-1)(e^{\beta_R\hbar\omega'}-1)},
\end{equation}
and 
\begin{equation}
\label{eq:Tdep2}
    f_b(\omega,\omega')=\frac{e^{\beta_L\hbar\omega'}(e^{\beta_L\hbar\omega}-1)}{(e^{\beta_L\hbar(\omega+\omega')}-1)(e^{\beta_L\hbar\omega'}-1)},
\end{equation}

From the three-phonon Landauer formula, we can reveal the following important findings. At the low temperature limit $\beta\rightarrow \infty$, this distribution function tends to $f\rightarrow 0$ as $e^{-\beta \omega}$, implying that the three-phonon scattering vanishes faster than the two-phonon process at low temperature. At the high temperature regime $\beta\rightarrow 0$, on the other side, the distribution function tends to infinity $f\rightarrow\infty$ as $1/\beta$ in the high-temperature limit, implying that in the weak coupling limit, the transmission function is unbounded and importantly, it increases linearly with temperature in the high temperature regime. This finding is in a stark contrast to the two-phonon scattering for which the phonon transmission is always bounded. 

Next we will discuss the temperature dependence of thermal conductance in the high temperature regime. In the high temperature limit, the Bose-Einstein distribution function is proportional to temperature, $n(\omega)\propto\frac{k_B T}{\hbar\omega}$. By using this relation, we find that the heat current becomes
\begin{equation}
        I_{h}=k_B(T_L-T_R)\int_0^\infty \frac{d\omega}{2\pi} [\bar{T}_{2p}(\omega)+\bar{T}_{3p}(\omega,T_L,T_R)].
\end{equation}
Therefore, its contribution to the thermal conductance $\sigma=I_{h}/(T_L-T_R)$ only depends on the transmission function. 

For the two-phonon processes, the transmission function remains unchanged as $\bar{T}_{2p}(\omega)=K_{i,j}K_{k,l}\Gamma_{ik}(\omega)\Gamma_{jl}(\omega)$ is a temperature-independent function. As a result, this thermal conductance is temperature-independent in the high-temperature limit.

However, for the three-phonon processes at the high temperature regime, we find that Eq.~(\ref{eq:Tdep1}) and Eq.~(\ref{eq:Tdep2}) can be reduced to
\begin{equation}
    f_a(\omega,\omega')=\frac{k_BT_R\omega}{\hbar\omega'(\omega+\omega')},
\end{equation}
and
\begin{equation}
    f_b(\omega,\omega')=\frac{k_BT_L\omega}{\hbar\omega'(\omega+\omega')}.
\end{equation}
As a result, the transmission function becomes linearly dependent on temperature
\begin{equation}
\bar{T}_{3p}(\omega,T_L,T_R)= T_L C_L(\omega)+ T_R C_R(\omega),
\end{equation}
where $C_L(\omega)$ and $C_R(\omega)$ are temperature independent quantities given by
\begin{equation}
    C_{L(R)}(\omega)=-{8k_B}\int_{-\infty}^{\infty}\frac{d\omega'}{2\pi}\frac{\omega S_{b(a)}(\omega,\omega')}{\omega'(\omega+\omega')}.
\end{equation}
The singularities in calculating $C_L$ and $C_R$ come from the phonons in long-wavelength limit and should be cancelled by zeros in the spectral densities from the scattering matrix $S$. However, in numerical calculation, convergence test is necessary to confirm a proper integration is performed.
This result is also consistent with the previous molecular dynamics simulation \cite{Le2017}, which shows that the thermal conductance for an anharmonic interface connecting two harmonic bulks will increase linearly with temperature in the Ar/heavy-Ar interface model. However, practically, such linearly increasing thermal conductance with temperature should be affected by the melting of the materials, or the higher order phonon-scattering scattering processes that may become dominating. Clearly, these processes will cause the breakdown of the linear trend.  

Interestingly, we also note that $\bar{T}_a$ only depends on $T_R$ while $\bar{T}_b$ only depends on $T_L$, which means the transmission probability due to the three-phonon scattering only depends on the temperature of the two-phonon side, but not the temperature of the one-phonon side. This feature remarkably explains the origin of the thermal rectification effect. For each scattering process, the temperature of the two-phonon side will change by reversing the temperature bias, which leads to a difference in transmission function and hence thermal conductance. So the rectification effect  essentially originates from the asymmetry of the temperature dependence in $\bar{T}_a$ and $\bar{T}_b$. This also indicates that rectification effect due to three-phonon scatterings will vanish in the limit $T_L\rightarrow T_R$. This finding also provides an important guideline for enhancing rectification effect: One needs to find two materials with distinct temperature-dependence of phonon properties, for instance, one has pulse-like sharp phonon density of state while the other has wide-spread phonon spectrum. 

\section{Application and numerical results}
Our above theoretical analysis has concluded that the thermal conduction of two-phonon processes should dominate at low-temperature regime while that of three-phonon processes should ultimately dominate at high temperature regime due to its unbounded nature. In general, such behavior should be highly material-dependent. Then an interesting question is: For representative systems, what is the role of three-phonon processes at room temperature?  Here we use graphene (Gr) and monolayer molybdenum disulfide (MoS$_2$) as examples to investigate the temperature-dependent behavior of three-phonon processes at the following three vdW interfaces: Gr-Gr, Gr-MoS$_2$ and MoS$_2$-MoS$_2$ interfaces. 

Within each layer, the phonon behaviors are studied using Quantum Espresso based on density functional theory (DFT) \cite{Giannozzi_2009}. 
The phonon spectral function is evaluated from the interatomic force constants obtained from the DFT calculations. We make the k-point mesh dense enough so that the spectral density obtained is convergent and continuous.
The interatomic force constants (IFCs) within the layer are obtained using DFT calculations, while the inter-layer IFCs are obtained through the Taylor expansion of inter-atomic Lennard-Jones potential $V(r)=4\varepsilon((\frac{r}{\sigma})^{12}-(\frac{r}{\sigma})^6)$, in order to ensure that second- and third-order inter-layer IFCs are obtained consistently. Here $r$ is the interatomic distance, $\varepsilon$ and $\sigma$  are energy and distance parameters depending on the species of atoms.  The used  parameters are $\varepsilon_{C-S}=7.35$meV, $\varepsilon_{C-Mo}=3.32$meV, $\sigma_{C-S}=3.513${\AA}, and $\sigma_{C-Mo}=3.075$\AA, which are obtained from universal force field model \cite{Rappe1992, Ding2016}. The inter-layer potentials up to 5 unit cells are considered. For the Gr-Gr calculation, the two layers are in AB stacking. For the MoS$_2$-MoS$_2$ calculation, the two layers are also in AB stacking, which has been demonstrated as the most stable stacking \cite{van_Baren_2019}. 
\begin{figure}
    \centering
    \includegraphics[width=\linewidth]{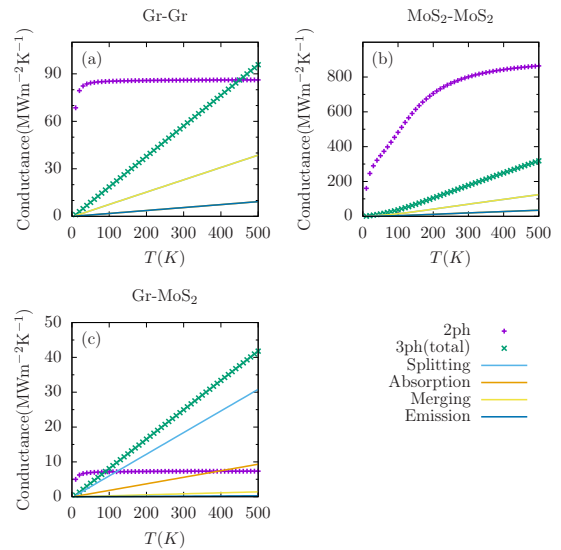}
    \caption{Temperature-dependence of thermal conductance contributed by two-phonon processes and three-phonon processes for (a) Gr-Gr interface, (b) MoS$_2$-MoS$_2$ interface and (c) Gr-MoS$_2$ interface.}
    \label{fig:Tdepend}
\end{figure}

The results of the calculations are shown in Fig.~\ref{fig:Tdepend}. For the Gr-Gr interface, we find that the contribution from two-phonon processes quickly saturates at around $50$K while that from three-phonon processes increases linearly with temperature, consistent with our theoretical finding. The contribution from three-phonon processes surpasses two-phonon processes at $T=460$K. At room temperature, three-phonon processes contribute 40.0\% of the total thermal conductance. Within the three-phonon scattering processes, the splitting and merging processes make the major contribution, in comparison with the minor contribution from the absorption and emission processes. Here we find that the contribution of phonon splitting to thermal conductance equals that of phonon merging and the contribution of phonon absorption equals that of phonon emission. This is because the interface is made up by the same material.

For the MoS$_2$-MoS$_2$ interface, we find that two-phonon conductance saturates at a relatively high temperature, above $500$K, while the three-phonon contribution increases slowly and does not surpass the two-phonon contribution even at $T=500$K. At room temperature, it only accounts for 18.1\% of the total thermal conductance. It is found that the splitting and merging processes are still the dominating three-phonon processes.

In contrast to Gr-Gr and MoS$_2$-MoS$_2$ interfaces, a distinctively different behavior is observed for the Gr-MoS$_2$ interface. Surprisingly, the three-phonon processes dominate the two-phonon processes in Gr-MoS$_2$ interface once the temperature surpasses as low as $T=90$K, far below room temperature. At room temperature, the three-phonon processes already account for 77.4\% of the total thermal conductance. This surprising observation can be explained from the following two aspects: (1) Due to the high-mismatch in the phonon spectra between graphene and MoS$_2$, the two-phonon processes are largely suppressed at Gr-MoS$_2$ interface, which is evidenced by noting that two-phonon contribution of Gr-MoS$_2$ interface  is less than both Gr-Gr interface and MoS$_2$-MoS$_2$ interface. (2) Due to the fact that the phonon frequency does not have to be conserved during the three-phonon scattering, the mismatch of the phonon spectra does not affect the three-phonon scatterings. In fact, it is the splitting process from Gr to MoS$_2$ that is dominating in the three-phonon processes.

\begin{figure}
    \centering
    \includegraphics[width=\linewidth]{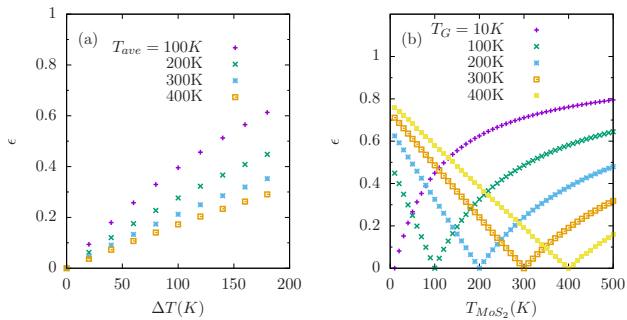}
    \caption{Thermal rectification due to three-phonon scattering. (a) The rectification coefficient is plotted against temperature bias. (b) The rectification coefficient is plotted against temperature of MoS$_2$ at various temperatures of graphene.}
    \label{fig:rect}
\end{figure}

Since experimentally one can only measure the total thermal conductance comprised of both harmonic and anharmonic contributions, so an interesting question is how to experimentally confirm the importance of three-phonon process. A simple approach is to measure the temperature dependence of interfacial thermal conductance. Our findings that the conductance due to two-phonon processes will saturate at high temperature and the conductance due to three-phonon processes is unbounded and increases linearly with temperature provide a clear rule for the differentiation. Besides such quantitative calibration of the relationship between temperature and interfacial thermal conductance, we also propose an easy approach to detect the anharmonicity in the interface scattering: The measurement of thermal rectification effect. We have shown that three-phonon processes are responsible for thermal rectification effect. Thus a pronounced thermal rectification effect is the signature of phonon anharmonicity. Here we use the Gr-MoS$_2$ interface for demonstration.  The result is shown in Fig.~\ref{fig:rect}. We find that the rectification coefficient (defined as $\epsilon=(G_h-G_l)/G_h$, where $G_h$ is the high thermal conductance, while $G_l$ is the low thermal conductance by reversing the temperature bias) increases linearly with increasing temperature bias (Fig.~\ref{fig:rect}(a)). The thermal rectification effect is more significant at low average temperature. For the average temperature of $300$K, $\epsilon$ reaches 0.1 at a temperature bias of $40$K. This indicates that the thermal current flowing from Gr to MoS$_2$ is around 90\% of the thermal current flowing from MoS$_2$ to Gr after reversing the temperature bias.

In Fig~\ref{fig:rect}(b), we plot the rectification coefficient against the temperature of MoS$_2$ at various temperatures of graphene. It is seen that the rectification coefficient decreases linearly with increasing temperature of MoS$_2$ and reaches 0 when $T_{M}=T_{G}$, and then increases again at high temperature regime. For example, $\epsilon$ reaches 0.2 when the temperature of MoS$_2$ is $400$K and the temperature of graphene is $300$K. The variations of recification coefficient with temperatures at Gr and MoS$_2$ reflect the phonon anharmonicity at the Gr and MoS$_2$ interface. 

\section{Conclusion}
We develop a quantum transport theory to deal with thermal conduction across a weakly interacting interface by placing harmonicity and anharmonicity under the same theoretical framework. For the anharmonic scattering, we focus on the dominating three-phonon processes, and find that the thermal conductance due to three-phonon processes can also be cast into Landauer formula albeit with the transmission function being temperature-dependent. However, the transmission function for the three-phonon processes decreases by $e^{-1/T}$ at low temperature limit, but increases \textit{linearly} with temperature at high temperature limit, suggesting that the anharmonic interface thermal conductance is unbounded at high temperature limit.  We also find that the two-phonon processes dominate the Gr-Gr and MoS$_2$-MoS$_2$ interfaces at room temperature, but three-phonon processes dominate the Gr-MoS$_2$ interface at a temperature as low as 90K, and at room temperature, the three-phonon processes account for 77.4\% of the total thermal conductance.
We also reveal that the physical origin of the thermal rectification effect and show that three-phonon processes can cause a measurable thermal rectification effect at the Gr-MoS$_2$ interface. The present work not only reveals important new sights into the thermal conduction across weakly interacting interfaces and but also provides useful guideline for thermal management and heat dissipation in electronic devices based on weakly-interacting materials.

\end{document}